\documentclass[aps,twocolumn,superscriptaddress,amsfont,graphicx,preprintnumbers]{revtex4-1}
\usepackage{subcaption,multirow}
\usepackage{amsmath,amssymb,graphicx,epsf,slashed,pstricks,hyperref,color}
\usepackage{dcolumn}
\usepackage{bm}
\usepackage{color}
\usepackage{multirow}
\usepackage{hhline}

 \newcommand{\lsim}{{\;\raise0.3ex\hbox{$<$\kern-0.75em\raise-1.1ex\hbox{$\sim$}}\;}}
\newcommand{\gsim}{{\;\raise0.3ex\hbox{$>$\kern-0.75em\raise-1.1ex\hbox{$\sim$}}\;}}
\newcommand{\beq}{\begin{equation}}
\newcommand{\eeq}{\end{equation}}
\newcommand{\bea}{\begin{aligned}}
\newcommand{\eea}{\end{aligned}}

\def\baa{\begin{array}}
\def\eaa{\end{array}}

\mathchardef\minus="002D

\newcommand{\amc}{{\sc MadGraph5\textunderscore}a{\sc MC@NLO}}
\newcommand{\delphes}{{\sc Delphes}}

\begin{document} 


\title{\boldmath Probing New Physics by the Tail of the Off-shell Higgs in $V_LV_L$ Mode
}

\preprint{CTPU-PTC-18-15} 

\author{Seung J. Lee}
\email{sjjlee@korea.edu}
\affiliation{Department of Physics, Korea University, Seoul 136-713, Korea}
\affiliation{School of Physics, Korea Institute for Advanced Study, Seoul 130-722, Korea}

\author{Myeonghun Park}
\email{parc.seoutech@seoutech.ac.kr}
\affiliation{Seoul National University of Science and Technology, Seoul, Korea}
\affiliation{Center for Theoretical Physics of the Universe, Institute for Basic Science (IBS), Daejeon, 34051, Korea}
\author{Zhuoni Qian}
\email{zhuoniq@ibs.re.kr}
\affiliation{Center for Theoretical Physics of the Universe, Institute for Basic Science (IBS), Daejeon, 34051, Korea}

\begin{abstract}

Off-shell Higgs at the high mass tail may shed light on the underlying mechanism of the electroweak symmetry breaking.
In the Standard Model, there is an exact cancellation of the logarithmic divergence between the box and Higgs-mediated triangle diagrams due to unitarity, such that the $gg\to ZZ (WW)$ process in the SM is dominated by the $V_T V_T$ transverse-mode. 
The cancellation can be delayed to a higher scale, when there is sufficiently large new physics contribution resulting in $V_LV_L$ longitudinal mode, which is commonly the case when the Higgs sector is modified. Thus the $V_LV_L$ final states in the high mass tail can be utilized as a sensitive probe for new physics. In this letter, we propose to utilize the information in angular observables to maximize the hint of a new physics hiding in the polarization of gauge bosons.
\end{abstract}
\maketitle

\section{Introduction}
\label{sec:description}

With the discovery of the Higgs boson at the LHC\,\cite{Aad:2012tfa, Chatrchyan:2012xdj}, much remains to be answered about the electroweak symmetry breaking (EWSB). 
For example, we do not know yet why the electroweak scale is so much lower than the planck scale, or how EWSB is triggered. These questions invoke the possibility of a Higgs sector being a portal to the physics beyond the Standard Model (BSM).
From the large amount of data being accumulated at the LHC, the LHC becomes a ``Higgs factory" at hand. Thus we can start our next journey by improving sensitivity over the general class of  a new physics (NP) which can modify the Higgs sector. 
Particularly, the off-shell Higgs deserves a careful study, optimally in the $h^*\to ZZ\to 4\ell$ decay channel for precision\,\cite{Englert:2014ffa, Azatov:2014jga, Goncalves:2018pkt}. 

In this letter we point out that the vector boson pair from the modified Higgs sector typically deviates in the longitudinal modes towards the high mass off-shell region of the Higgs.
Closely related to EWSB, the importance of probing the longitudinal component of the vector bosons have long been discussed\,\cite{Han:2009em}. The relevance of the longitudinal mode at high mass scale is expected by the Goldstone boson equivalence theorem and it is natural to consider tagging the longitudinal polarization mode for improving our understanding of the Higgs sector.  

Depending on the different polarization combinations, the vector boson pair production could be categorized into TT (transverse-transverse), TL(transverse-longitudinal), and LL(longitudinal-longitudinal) modes. 
In the Standard Model (SM), $Z$ boson pair production is mostly from $q\bar{q}\to Z Z$ process, which is dominated by TT mode. For $gg\to Z Z$, the destructive interference between the massive quark box-loops and triangle-Higgs diagrams leaves the total cross section dominated by TT mode.
To understand the cancelation in LL mode, we note that the box diagram with massive quark contribution scales as $(\sqrt{\hat s}/m_Z)^2$ in axial-axial current with a coupling $C_A$ and the corresponding amplitute in the $\sqrt{\hat s} \gg m_t$ limit goes\,\cite{Baur:1988cq},
\begin{equation*}
 \mathcal{A}_{\Box} \to -8 C_A^2\frac{m_t^2}{\hat s}\left( \frac{\hat s}{m_Z^2}\right)\log ^2\left(\frac{\hat s}{m_t^2}\right)  = -\frac{m_t^2}{2m_Z^2}\log ^2\left(\frac{\hat s}{m_t^2}\right).
\end{equation*}
At high energy, this amplitude violates unitarity, and so is the amplitude from the Higgs diagram, 
\begin{equation*}
\mathcal{A}_{\bigtriangleup}\to \frac{m_t^2}{\hat s} \frac{1}{2} \log^2\left(\frac{m_t^2}{\hat s}\right) \left(\frac{\sqrt{\hat s}}{m_Z}\right)^2  = \frac{m_t^2}{2m_Z^2}\log ^2\left(\frac{\hat s}{m_t^2}\right).
\end{equation*}
The $\log$-divergent terms from the box and Higgs contribution cancel exactly, and unitarity is restored.
However, this exact cancellation may be delayed in the presence of NP at a scale probable by the LHC. In this case, a modification of $gg \to V_L V_L$ amplitude can be established from 1) a change in Higgs propagator 2) introduction of a new propagator 3) a variation in $hV_LV_L$ form factor.  These cases can be portrayed with following examples. Higgs portal with a light scalar loop corrects the Higgs self energy and modifies the Higgs propagator correspondingly. A heavy Higgs-like scalar with broad-width contributes to $gg\to V_L V_L$ as with an additional propagator. Finally in the case of quantum critical Higgs (QCH) model, where Higgs arises as a bound state of conformal field theory, deviation rises up above the continuum scale, while unitarity is restored at much higher scale\,\cite{Bellazzini:2015cgj}.
Tagging the LL mode in the tail of off-shell Higgs provides a sensitive probe to all these NP scenarios.
In a case when the NP is above the LHC search scale and its effect could be integrated out, we  consider higher dimensional effective field theory (EFT) operator. For the study of LL mode, the contributing gauge-invariant EFT operator starts from dimension eight\,\cite{Bellazzini:2018paj},
\begin{equation} 
O^{(8)}=\frac{c_8}{\Lambda^4}\left( i\bar{\psi}^{\{\mu}\partial^{\nu\}} \psi+\rm{h.c.} \right ) D_\mu H^\dagger D_\nu H.
\end{equation}

Experimental searches exist on specific cases of NP in the Higgs sector, through $ZZ$ final states, making full use of the kinematics in the final states for heavy resonance searches\,\cite{Sirunyan:2018qlb, Aaboud:2017rel}, and indirect bounds on Higgs total width from off-shell Higgs signal measurement\,\cite{Kauer:2012hd, Khachatryan:2014iha, Aaboud:2018puo}. 
For a general search to find hints in a Higgs sector, however, we cannot rely on those dedicated analyses which require model-specific information on a NP model. 
We show a strategy for a general analysis to improve sensitivity on three distinct NP cases and the EFT operator.

\section{New physics within the LHC scale}
\label{sec:examples}
\subsection{Higgs portal light scalar}

As studied in\,\cite{Goncalves:2017iub}, a Higgs portal light scalar with mass of $m_S>m_h/2$, which evades constraints from Higgs invisible decay searches, would contribute through loop effects to the Higgs self energy, and modify the $m_{ZZ}$ distribution at high energy scale. Such a light scalar that couples only through the Higgs without mixing is otherwise poorly constrained except from the precision measurements of $Zh$ inclusive production at a future lepton collider\,\cite{Asner:2013psa, Mangano:2019rww, An:2018dwb}. To study, we write a simplified Lagrangian of a SM plus a complex scalar in the form:
\beq
\mathcal{L} = \mathcal{L}_{\rm SM} + \partial_\mu S\partial^\mu S^* - \mu^2|S|^2 - \kappa |S|^2|\Phi|^2.
\eeq
At Next Leading Order (NLO), the scalar $S$ modifies the Higgs propagator through one loop, and the 
renormalized self-energy becomes,
\beq
\hat\Sigma_h(s) = \Sigma_h(p^2) - \delta \mu_h^2 + (p^2 - \mu_h^2)\delta Z_h.
\eeq
$\mu_h^2$ is the square of complex mass defined as $\mu_h^2 = m_h^2 - i m_h \Gamma_h$. $\delta Z_h$ and $\delta \mu_h^2$ are the wave function and mass renormalization of the Higgs field, respectively. In the on-shell scheme, they are defined as,
\beq
\delta \mu_h^2 = \Sigma_h (\mu_h^2),~~ \delta Z_h = -\frac{d\Sigma_h}{d p^2}(\mu_h^2).
\eeq
Exact NLO precision is calculated by truncating to order $\mathcal{O}[(\hat\Sigma_h(p^2))^2]$ at squared amplitude level, with the modified propagator
\begin{equation}
{\rm Propagator}^{\rm NLO}  = \frac{i}{p^2 - \mu_h^2}\left(1 - \frac{\hat\Sigma_h(p^2)}{p^2 - \mu_h^2}\right).
\end{equation}

To understand the cross section deviation with modified propagator, we see that when $\sqrt{\hat s}>2m_S$ the renormalized self energy $\hat\Sigma_h(p^2)$ are complex. $\rm{Im}(\hat\Sigma_h)$ corresponds to the decay amplitude of $h\to SS$, and turns on at $\sqrt{\hat s}$ above $2 m_S$. The Re part $\rm{Re}(\hat\Sigma_h)$ changes the magnitude of the amplitude of the Higgs contribution. 
This deviation at high energy becomes apparent in the LL mode as expected. 
We set the benchmark point with $\kappa = 9$ and $m_S = 80$ GeV, where $\kappa$ is set to be largest parameter still allowed by current experimental search\,\cite{Sirunyan:2018qlb, Kang:2018jem}.

\begin{figure*}[!t]
\includegraphics[width=0.4\linewidth]{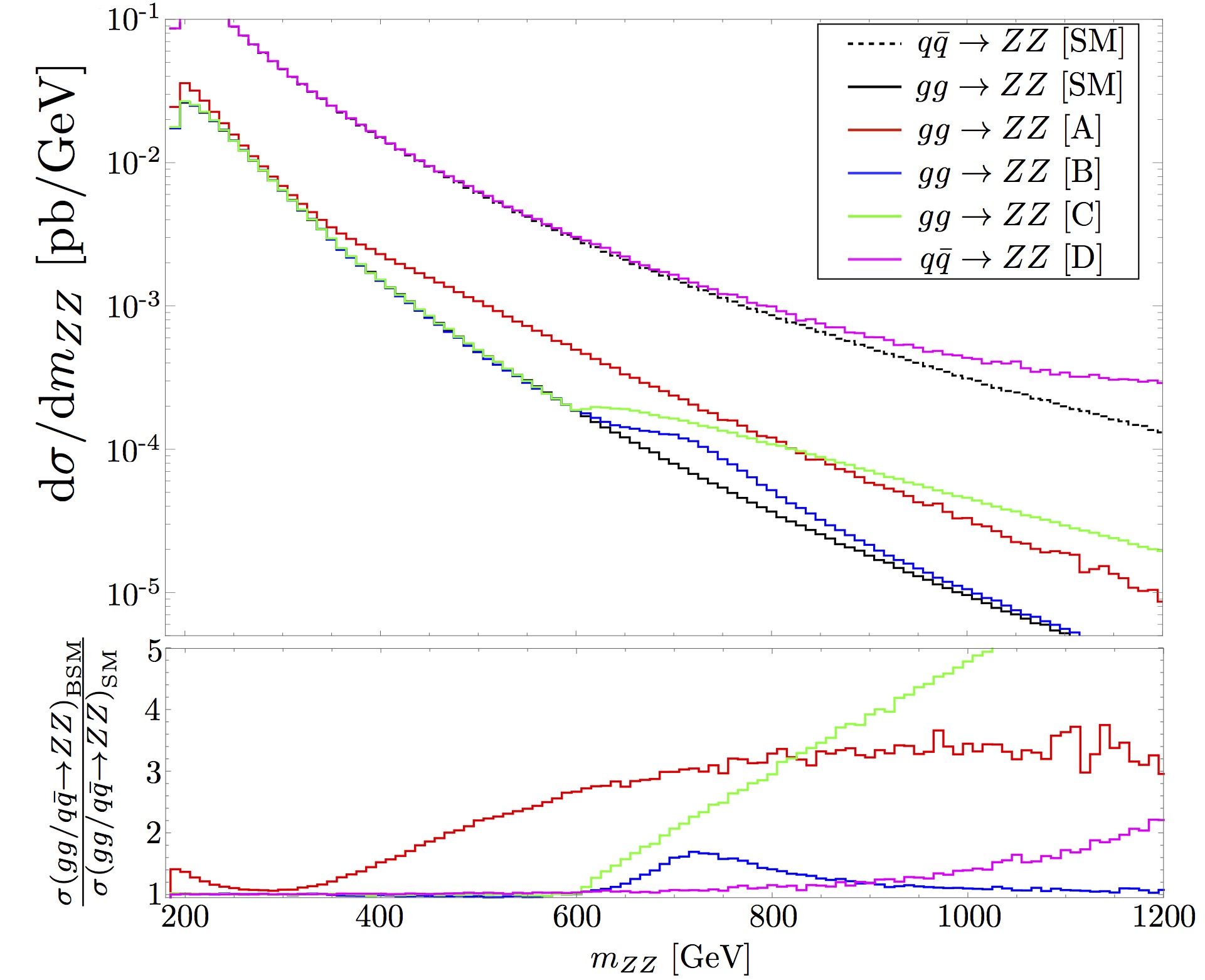}\quad
\includegraphics[width=0.4\linewidth]{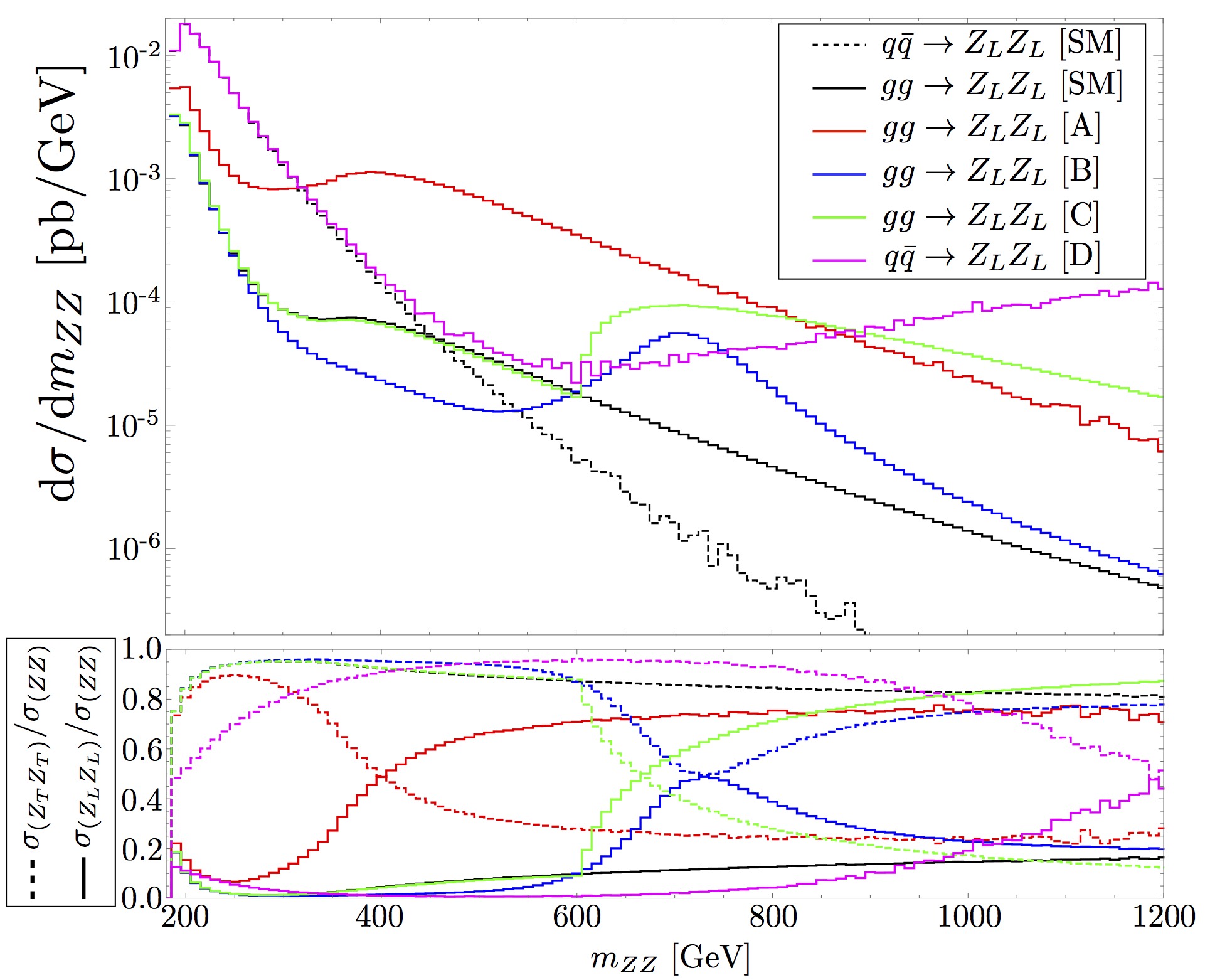}
\caption{(Left top) The differential cross sections of $q\bar q\to ZZ$, $gg\to ZZ$ for relevant processes, and the ratio (left bottom) between the BSM cases A, B, C, D and the SM backgrounds. (Right top) The differential cross section in the LL polarization mode for relevant processes, and the ratio (right bottom) of LL and TT mode components compared to the total rate for $gg(q\bar q)\to ZZ$ processes.}
	\label{fig:caseabc_pol}
\end{figure*}

\subsection{Broad-width heavy scalar}

Another example is a heavy scalar $S$ that decays to $ZZ$, with amplitude proportional to its mixing with the SM scalar doublet. 
Here we take a representative example of an additional real scalar as,
\begin{equation}
\mathcal{L} \ni \mathcal{L}_{\rm SM} - \mu_S S|\Phi|^2\,.
\end{equation}
After EWSB, there is mixing between the $S$ and the Higgs $h$ with mixing angle $\tan\theta = \frac{\mu_S v}{\sqrt{(\mu_S v)^2 + (m_S^2- m_h^2)^2}}$, here $v$ is the SM vacuum expectation value. In the limit $m_S^2 \gg m_h^2$ and small mixing, $\sin\theta \sim \mu_S v / m_S^2$\,\cite{DiLuzio:2016sur}.
Through mixing, all the Higgs couplings to the other SM particles are rescaled by $\cos\theta$, while the $SXX$ couplings are $\sin\theta$ times the SM Higgs coupling value.
We take the scalar mass at $M_S = 700$ GeV and $\sin\theta = 0.4$ as still allowed by current Higgs data\,\cite{Tao:2018zeu}, and assume a relatively broad width $\Gamma_S = 140$ GeV which manifests large interference. This corresponds to $\mu_S\sim 4.6\, v$ and the decay width of the heavy scalar to SM around 30 GeV, dominated by $S\to hh$. The broad width we assume could arise from the real scalar $S$ decaying to Hidden sector, which could eventually decay back to soft SM final states. This case is constrained by mono-jet searches\,\cite{Falkowski:2015swt}. 

\subsection{Quantum Critical Higgs}

Quantum critical Higgs type of models\,\cite{Stancato:2008mp, Falkowski:2008fz, Falkowski:2008yr, Bellazzini:2015cgj} typically predict a higher scale continuum which modify the Higgs off-shell region. The natural version of quantum critical Higgs can be built by implementing the original model in the warped extra dimension with linear dilaton set-up, as in the continuum naturalness framework\,\cite{Csaki:2018kxb}, except that the Higgs here would be represented by generic bulk scalar field, instead of 5th component of a bulk gauge field corresponding to 4D pseudo-Nambu-Goldstone boson. 
In general, the Higgs couplings to other SM particles could depend on the details of the UV theory, the conformal symmetry breaking and the scale. We consider a minimal scenario where the propagator of the physical Higgs field and the $hZZ$ coupling are modified as follows, 
\begin{eqnarray}
G_h(p) &&= -  \frac{i Z_h}{(\mu^2-p^2-i\epsilon)^{2-\Delta}-(\mu^2-m_h^2)^{2-\Delta}}, \nonumber \\
g_{hZZ} &&= -  \frac{(\mu^2)^{2-\Delta}-(\mu^2-p^2)^{2-\Delta}}{\hat s} g_{hZZ}^{\rm SM}.
\end{eqnarray}
The non-standard $hZZ$ form factor arises from gauge invariant form of the Higgs two-point function. The continuum scale $\mu$ and the anomalous dimension $\Delta$ are the two new parameters in the simplified case. We chose $\mu=600$ GeV and $\Delta=1.6$ as benchmark point, which is set to be still allowed by current experimental search in the high mass tail of $ZZ$ final states\,\cite{Sirunyan:2018qlb}.

\section{Analysis}
\label{sec:analysis}
Focusing on the polarization composition of  above NP cases,
from Fig.\,\ref{fig:caseabc_pol}, we see the deviation of BSM compared to the SM in the total rate (Left plot) is mainly from deviation in the LL mode (Right plot). 
Given this physical feature, we enhance the experimental sensitivity on generic Higgs sector new physics, that shows up in the high energy scale. 

We use \amc\,\cite{Alwall:2014hca} to generate $gg (q\bar q)\to ZZ \to e^- e^+ \mu^- \mu^+$ events at QCD leading order, and rescale with a k-factor of 1.8 (1.5) for the $gg(q\bar q)$ initiated processes respectively\,\cite{Caola:2015psa, Cascioli:2014yka}. For NP examples, we generate $gg \to ZZ \to e^- e^+ \mu^- \mu^+$ events for the light scalar case (case A), heavy Higgs case (case B) and QCH model (case C) with model parameters given in Sec.~\ref{sec:examples}. Generator level cuts are applied as $p_{T \ell} > 10$ GeV, $|\eta_{\ell}| < 2.5$, $m_{\ell^- \ell^+} > 50$ GeV, and $m_{4\ell} > 560$ GeV. After detector simulation with \delphes\,\cite{deFavereau:2013fsa}, we further require in the final state a pair of electrons and muons with basic cuts,
\begin{equation}
80<m_{ll}<100~ {\rm GeV},~~ m_{4l}>600~{\rm GeV}.
\label{eqn:bcut}
\end{equation}

\begin{figure*}[!t]
\centering
	\begin{minipage}{.3\textwidth}
		\includegraphics[width=1.05\linewidth]{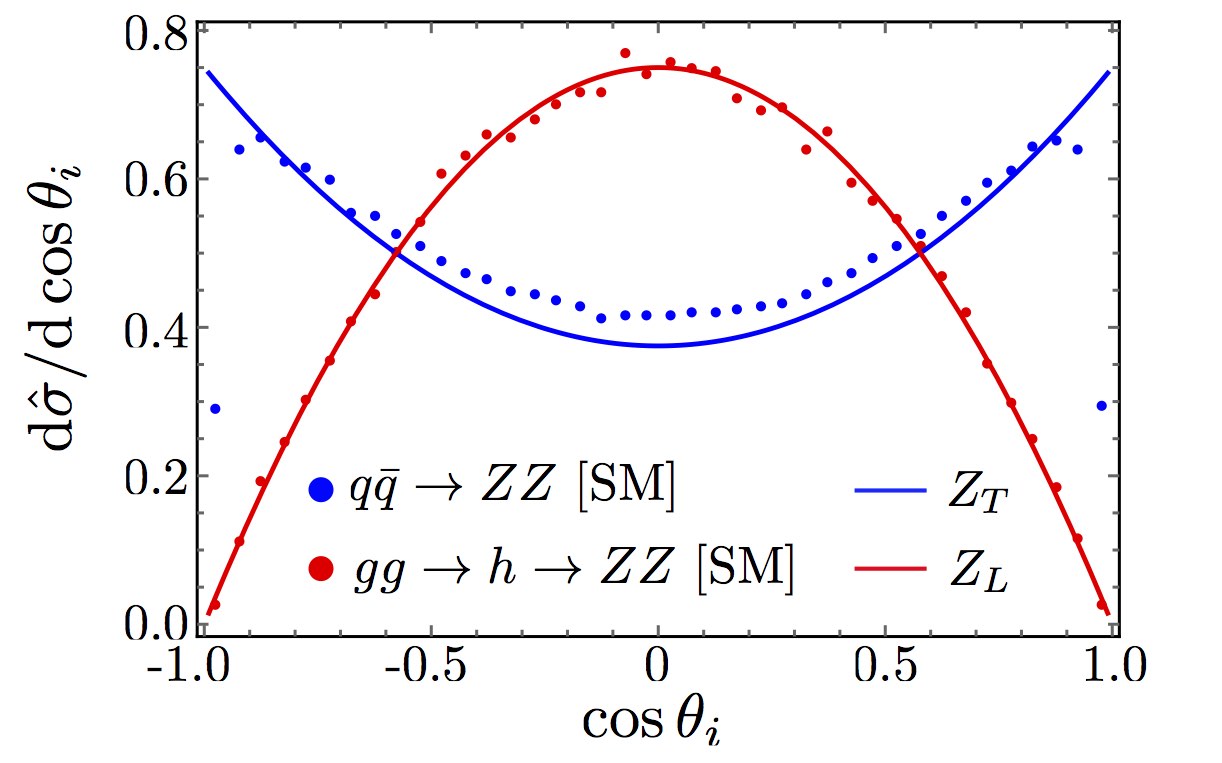}
	\end{minipage}%
	\begin{minipage}{.3\textwidth}
		\includegraphics[width=1.05\linewidth]{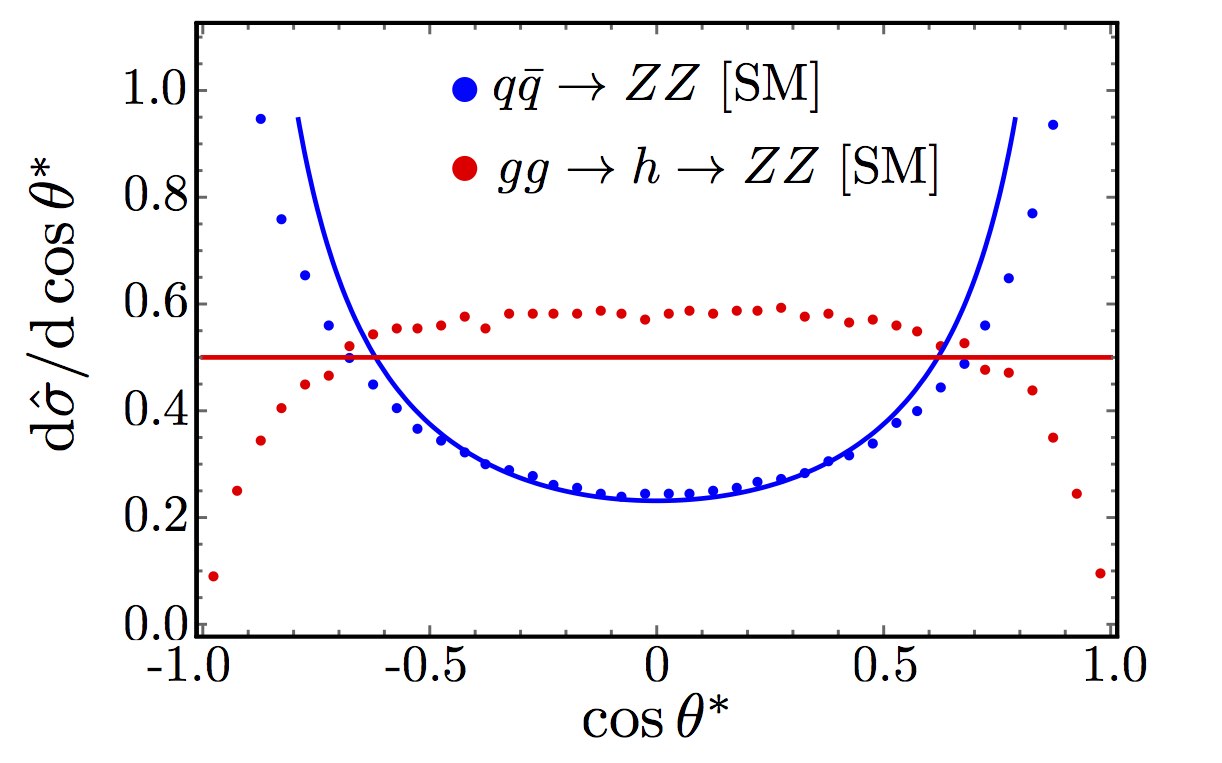}
	\end{minipage}
	\begin{minipage}{.30\textwidth}
		\includegraphics[width=1.0\linewidth]{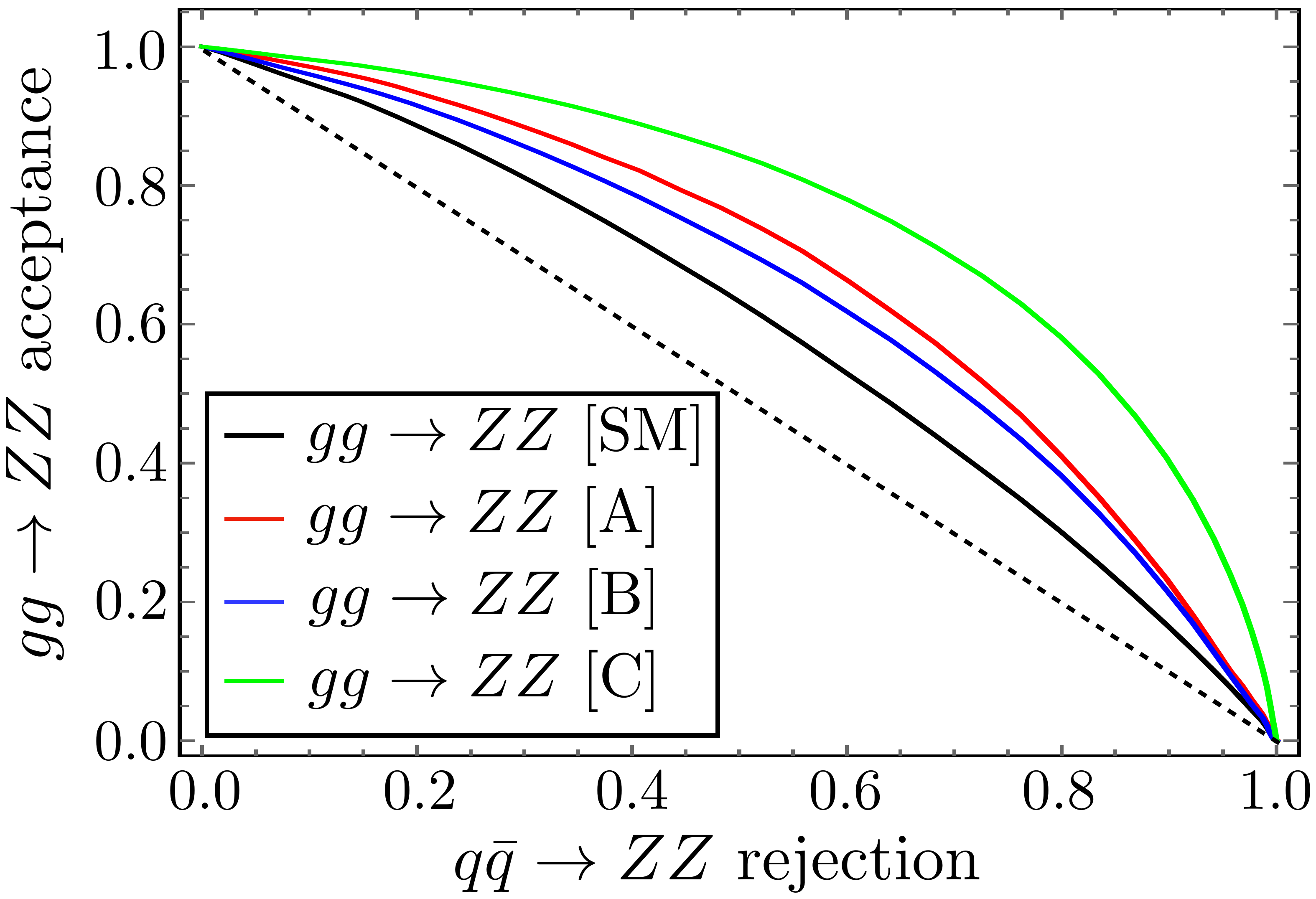}
	\end{minipage}%
	\caption{(Left) $\cos{\theta_1}$ distribution for signal and background. (Middle)  $\cos{\theta}^*$ distribution for signal and background.  Dotted lines are from Monte Carlo, and solid curves are from the theory prediction without cuts. (Right) ROC showing acceptance rate of the $gg$-initiated processes against the rejection rate of the SM $q\bar q$-initiated one.}
	\label{fig:angles}
\end{figure*}

The generic NP signal is $(gg\to h^*\to 4\ell)$-like, dominated by LL mode. We set  $m_h=m_{4\ell}$ to remove Higgs mass dependence in our analyses. 
Backgrounds are the SM $q\bar q\to ZZ$, and $gg\to ZZ$ processes, dominated by TT mode in the high mass scale.

One variable we find useful is $\cos\theta^*$. It's the cosine of the angle between one final state $Z$ that is reconstructed from $\mu^- \mu^+$ pair, with respective to the beam line at the center of mass frame. The $\cos\theta^*$ distribution for signal is flat, as expected from a s-channel scalar mediator. For the major background $q \bar q \to Z Z$, the contribution is mostly from t/u-channel diagrams,
\begin{equation}
\frac{\textrm{d}\sigma_{(q\bar q \to Z Z)}}{\textrm{d} \cos\theta^*}  \simeq \frac{\textrm{d}\sigma_{(q\bar q \to Z_T Z_T)}}{\textrm{d} \cos\theta^*}  \propto \frac{1+\cos^2\theta^*}{1-\cos^2\theta^*}
\end{equation}
in the limit  $\sqrt{\hat s} \gg m_Z$, which is a good approximation at the energy scale we are interested in. 
In the left plot of Fig.\,\ref{fig:angles}, we show the normalized $\cos\theta^*$ distribution from simulated events (dots) as well as the theoretical prediction without cuts (solid curve) for both the dominant background $q\bar q\to e^- e^+ \mu^- \mu^+$, and the signal $gg\to h^*\to e^- e^+ \mu^- \mu^+$. The agreement worsens at the edges due to detector efficiencies, while the qualitative features remain. We choose the optimized cut on $\cos\theta^*$ at the maximum of $S/\sqrt{B}$. Both simulation and analytic expression agrees at a cut around $|\cos\theta^*| < 0.7$.
We then apply a polar angle cut on the leading $Z_1$ decay of $|\cos\theta_1| < 0.68$, whose theoretical and simulated-event distributions are shown in the middle plot of Fig.~\ref{fig:angles}. Our angular cuts are defined as,
\begin{equation}
|\cos\theta^*| < 0.7,\quad|\cos\theta_1| < 0.68.
\label{eqn:acut}
\end{equation}

To further suppress the $q \bar q$-initiated background and maximize the sensitivity to new physics, we separately perform a Boosted Decision Tree (BDT) analysis. First we feed BDT analysis with 6 variables: $m_{4\ell}$, $\cos\theta^*$, $\cos\theta_1$, $\cos\theta_2$, $\Delta\phi$ and KD. $\Delta\phi$ is the angle between the two decay planes of the $Z$ bosons. KD is defined as,
\begin{equation}
\textrm{KD} = 
\ln \left( 
\frac{f_g(x_1) f_g (x_2) |\mathcal{M}(gg\to h^* \to 4\ell)|^2}
{\sum f_{q}(x_1) f_{\bar q}(x_2) |\mathcal{M}(q\bar q \to 4\ell)|^2} \right),
\end{equation}
which is the ratio of the squared amplitudes between the signal and background processes, weighted with the respective PDFs\,\cite{Avery:2012um}. 
We use an adaptive boosting algorithm with 850 number of decision trees, and maximum depth of 3. Most important variables under BDT training are KD, $\cos\theta^*$ and $\cos\theta_1$. 
As defined above, the signal (SIG) for optimization is $gg\to h^*\to 4\ell$, where $m_{h^*} = m_{4\ell}^{\textrm{obs}}$ is set in the calculation of KD. Background (BKG) is the SM $q\bar q\to 4\ell$. 
The BDT classifier, optimized upon SIG and BKG, further reduces the $q \bar q$-initiated background.
BDT score is then calculated for all processes including the SM and the new physics cases considered. 
We show the ROC-curve from the BDT analysis in the right plot of Fig.\,\ref{fig:angles}. The $y$-axis shows the acceptance of the different $gg$-initiated processes including the SM, case A, B and C, the $x$-axis is the corresponding rejection rate of the $q\bar q$-initiated SM background.
Eventually, in Table\,\ref{tab:res} we list the significance achieved at the basic cuts, and then a angular cuts or BDT cut, 
respectively. 
We also show the luminosity needed to achieve a 3$\sigma$ significance for each NP case.

\begin{table}[!tb]
\centering 
\begin{tabular}{|c|c|c|c|c|c|}
 \hline
 Significance $\sigma$ & case A & case B & case C \\ \hline
with basic cuts & 2.01  & 0.634 & 4.71  \\ \hline
with basic + angle cuts  & 2.32  & 0.838 & 5.78  \\ \hline
with basic cuts + BDT & 2.45  & 0.92 & 7.01 \\\hhline{|=|=|=|=|} 
\rule{0pt}{2.5ex}    Luminosity  for 3$\sigma$ discovery & $4.2\textrm{ab}^{-1}$ & $29\textrm{ab}^{-1}$ & $0.5\textrm{ab}^{-1} $\\ \hline
\end{tabular}
\caption[]{Achievable sensitivities for the NP cases at 3 ab$^{-1}$ HL-LHC with benchmark points.
Basic cuts are defined in eq.\,(\ref{eqn:bcut}) and angle cuts in eq.\,(\ref{eqn:acut})}
\label{tab:res}
\end{table}

\section{Dimension-8 EFT Operator}
\label{sec:eft}
The dimension-8 operator gives rise to an energy dependent $q\bar q Z_L Z_L$ 4-point interaction. The relevant term for our $pp\to ZZ$ process, constrained by unitarity/analyticity\,\cite{Adams:2006sv, Bellazzini:2016xrt} reads~\cite{Bellazzini:2018paj},
\begin{equation}
-\frac{c_8}{\Lambda^4}\frac{i g_Z^2 v^2}{32}(\bar\psi_q \gamma^\mu\partial^\nu\psi_q + \bar\psi_q \gamma^\nu\partial^\mu\psi_q + h.c.) Z_\mu Z\nu.
\end{equation}
With benchmark point $\Lambda = 1.26$ TeV, and $c_8 = 4\pi$, the deviation is shown in Fig.\,\ref{fig:caseabc_pol}, mainly in the LL mode as expected. The deviation arises from interference between the dimension-8 operator and the SM $q\bar q\to ZZ$. Despite the smallness of the $ Z_L Z_L$ components in the SM, the interference becomes sizable at high mass tail due to the amplitude level order $\hat s^2$ enhancement. For validity of the EFT approach, we require $m_{4\ell}<1.2$ TeV in the analysis.

This scenario with adding a dimension-8 EFT operator is however different from the scenario A,B,C. Specifically, the $\cos\theta^*$ distribution is proportional to $\sin^2(2\theta^*)$ with a distinct di-peak structure. 
An angle cut of $|\cos\theta_1| < 0.68$ to favor the LL mode finds a 7\% improvement on the sensitivity. We then adopt a BDT analysis over variables $\theta^*,\theta_1,\theta_2, \phi, m_{4\ell}$, and found the $m_{4\ell}$ variable most important in discrimination, followed by $\cos\theta^*$ and $\cos\theta_{i}$. This is expected due to the sensitive energy dependence of the dimension-8 operator. We find that with 100 fb$^{-1}$  data of the 13\,TeV LHC, a cut of $\cos\theta_1 < 0.68$ constrains this operator to $\Lambda > 1.3 $ TeV at $1\sigma$, and BDT analysis improves the bound  to $1.5$ TeV.

As studied in the Ref.\,\cite{Bellazzini:2018paj}, there are in general $Z_T Z_T$ channels that get enhanced as well. The reach to these TT-mode enhancing operators is comparable to the LL-mode at around $\Lambda > 1.3 $ TeV, due to interference with a larger SM TT-mode but weaker center-of-mass energy enhancement. 

\section{Summary and Discussion}
\label{sec:conclusion}
We study the the high mass tail region of the $pp \to ZZ \to 4\ell$ channel for a precision, which is sensitive to the modification of the Higgs sector.
We point out that the deviation is mostly from the LL-mode of $Z$ bosons. The sensitivity in the LHC analysis would be improved by suppressing TT-mode 
with utilizing angular correlations.
To cover a case where a NP is within the LHC search scale, we evaluate three different NP scenarios which affect the higgs mediated diagrams in $gg\to ZZ$. Despite the different energy dependence in each NP, angular cuts favoring the central and LL mode would improve the sensitivity to probe all these scenarios by $20\%-30\%$. A BDT analysis optimized for LL mode that includes the the energy dependence and angular variables as well, would further improve all the sensitivities by about 20\%. If any of these Higgs sector new dynamics were around the corner for discovery, the High Luminosity (HL)-LHC would likely to observe a definite deviation in the tail region.  

We also examine the dimension-8 EFT operator which could arise to importance over the dimension-6 ones favored by certain symmetry models\,\cite{Bellazzini:2018paj}. The particular operator we study enhances the LL mode, but shows different distribution in the other variables such as $\cos\theta^*$ compared to the Higgs sector NP scenarios we consider. A cut of $\cos\theta_1 < 0.68$ which favors the longitudinal mode alone improves the current sensitivity by about $7\%$. 

We would like to point out that, a combined search with all the di-boson and Higgs associated production channels would accumulatively improve the sensitivity for  NP at the tail of off-shell Higgs. The study of these additional channels could be extended from our analysis. 
Once deviations in the high mass tail of $V_LV_L(h)$ are identified at HL-LHC, the next step is to discriminate between the NP scenarios. The different energy dependences as shown in Fig.\,\ref{fig:caseabc_pol} is one remaining variable, which could be employed to favor one scenario over another.

\begin{acknowledgments}
The authors are grateful for conversations with Tao Han.  This work was supported by the National Research Foundation of Korea (NRF) grant funded by the Korea government (MEST) (No. NRF-2015R1A2A1A15052408). SL was supported by Samsung Science and Technology Foundation under Project Number SSTF-BA1601-07.
MP is supported by Basic Science Research Program through the National Research Foundation of Korea Research Grant NRF-2018R1C1B6006572.  MP and ZQ are supported by IBS under the project code, IBS-R018-D1.
\end{acknowledgments}


\begin{thebibliography}{99}

\bibitem{Aad:2012tfa} 
  G.~Aad {\it et al.} [ATLAS Collaboration],
  Phys.\ Lett.\ B {\bf 716}, 1 (2012)
  doi:10.1016/j.physletb.2012.08.020
  [arXiv:1207.7214 [hep-ex]].

\bibitem{Chatrchyan:2012xdj} 
  S.~Chatrchyan {\it et al.} [CMS Collaboration],
  Phys.\ Lett.\ B {\bf 716}, 30 (2012)
  doi:10.1016/j.physletb.2012.08.021
  [arXiv:1207.7235 [hep-ex]].

\bibitem{Englert:2014ffa} 
  C.~Englert, Y.~Soreq and M.~Spannowsky,
  JHEP {\bf 1505}, 145 (2015)
  doi:10.1007/JHEP05(2015)145
  [arXiv:1410.5440 [hep-ph]].

\bibitem{Azatov:2014jga} 
  A.~Azatov, C.~Grojean, A.~Paul and E.~Salvioni,
  Zh.\ Eksp.\ Teor.\ Fiz.\  {\bf 147}, 410 (2015)
  [J.\ Exp.\ Theor.\ Phys.\  {\bf 120}, 354 (2015)]
  doi:10.1134/S1063776115030140, 10.7868/S0044451015030039
  [arXiv:1406.6338 [hep-ph]].
  
\bibitem{Goncalves:2018pkt} 
  D.~Goncalves, T.~Han and S.~Mukhopadhyay,
  Phys.\ Rev.\ D {\bf 98}, no. 1, 015023 (2018)
  doi:10.1103/PhysRevD.98.015023
  [arXiv:1803.09751 [hep-ph]].

\bibitem{Han:2009em} 
  T.~Han, D.~Krohn, L.~T.~Wang and W.~Zhu,
  JHEP {\bf 1003}, 082 (2010)
  doi:10.1007/JHEP03(2010)082
  [arXiv:0911.3656 [hep-ph]].

\bibitem{Baur:1988cq} 
  U.~Baur, E.~W.~N.~Glover and J.~J.~van der Bij,
  Nucl.\ Phys.\ B {\bf 318}, 106 (1989).
  doi:10.1016/0550-3213(89)90049-7

\bibitem{Bellazzini:2018paj} 
  B.~Bellazzini and F.~Riva,
  arXiv:1806.09640 [hep-ph].


\bibitem{Adams:2006sv} 
  A.~Adams, N.~Arkani-Hamed, S.~Dubovsky, A.~Nicolis and R.~Rattazzi,
  JHEP {\bf 0610}, 014 (2006)
  doi:10.1088/1126-6708/2006/10/014
  [hep-th/0602178].
  
\bibitem{Bellazzini:2016xrt} 
  B.~Bellazzini,
  JHEP {\bf 1702}, 034 (2017)
  doi:10.1007/JHEP02(2017)034
  [arXiv:1605.06111 [hep-th]].

  
\bibitem{Franceschini:2017xkh} 
  R.~Franceschini, G.~Panico, A.~Pomarol, F.~Riva and A.~Wulzer,
  JHEP {\bf 1802}, 111 (2018)
  doi:10.1007/JHEP02(2018)111
  [arXiv:1712.01310 [hep-ph]].


\bibitem{Goncalves:2017iub} 
  D.~Goncalves, T.~Han and S.~Mukhopadhyay,
  Phys.\ Rev.\ Lett.\  {\bf 120}, no. 11, 111801 (2018)
  Erratum: [Phys.\ Rev.\ Lett.\  {\bf 121}, no. 7, 079902 (2018)]
  doi:10.1103/PhysRevLett.120.111801, 10.1103/PhysRevLett.121.079902
  [arXiv:1710.02149 [hep-ph]].
  
  \bibitem{Asner:2013psa} 
  D.~M.~Asner {\it et al.},
  arXiv:1310.0763 [hep-ph].
  
\bibitem{Mangano:2019rww} 
  M.~L.~Mangano,
  doi:10.1142/9789813238053\_0017
    
\bibitem{An:2018dwb} 
  F.~An {\it et al.},
  arXiv:1810.09037 [hep-ex].

\bibitem{Sirunyan:2018qlb} 
  A.~M.~Sirunyan {\it et al.} [CMS Collaboration],
  JHEP {\bf 1806}, 127 (2018)
  doi:10.1007/JHEP06(2018)127
  [arXiv:1804.01939 [hep-ex]].
  
\bibitem{DiLuzio:2016sur} 
  L.~Di Luzio, J.~F.~Kamenik and M.~Nardecchia,
  Eur.\ Phys.\ J.\ C {\bf 77}, no. 1, 30 (2017)
  doi:10.1140/epjc/s10052-017-4594-2
  [arXiv:1604.05746 [hep-ph]].
  
\bibitem{Tao:2018zeu} 
  J.~Tao [CMS Collaboration],
  [arXiv:1810.00256 [hep-ex]].
  
\bibitem{Falkowski:2015swt} 
  A.~Falkowski, O.~Slone and T.~Volansky,
  JHEP {\bf 1602}, 152 (2016)
  doi:10.1007/JHEP02(2016)152
  [arXiv:1512.05777 [hep-ph]].
  
\bibitem{Bellazzini:2015cgj} 
  B.~Bellazzini, C.~Cs{\^a}ki, J.~Hubisz, S.~J.~Lee, J.~Serra and J.~Terning,
  Phys.\ Rev.\ X {\bf 6}, no. 4, 041050 (2016)
  doi:10.1103/PhysRevX.6.041050
  [arXiv:1511.08218 [hep-ph]].

\bibitem{Stancato:2008mp}
D.~Stancato and J.~Terning, \emph{{The Unhiggs}},
  \href{http://dx.doi.org/10.1088/1126-6708/2009/11/101}{\emph{JHEP} {\bf 11}
  (2009) 101}, [\href{http://arxiv.org/abs/0807.3961}{{\tt 0807.3961}}].

\bibitem{Falkowski:2008fz}
A.~Falkowski and M.~Perez-Victoria, \emph{{Electroweak Breaking on a Soft
  Wall}}, \href{http://dx.doi.org/10.1088/1126-6708/2008/12/107}{\emph{JHEP}
  {\bf 12} (2008) 107}, [\href{http://arxiv.org/abs/0806.1737}{{\tt
  0806.1737}}].

\bibitem{Falkowski:2008yr} 
  A.~Falkowski and M.~Perez-Victoria,
  Phys.\ Rev.\ D {\bf 79}, 035005 (2009)
  doi:10.1103/PhysRevD.79.035005
  [arXiv:0810.4940 [hep-ph]].

\bibitem{Csaki:2018kxb} 
  C.~Cs{\''a}ki, G.~Lee, S.~J.~Lee, S.~Lombardo and O.~Telem,
  arXiv:1811.06019 [hep-ph].
  


\bibitem{Sirunyan:2018qlb} 
  A.~M.~Sirunyan {\it et al.} [CMS Collaboration],
  JHEP {\bf 1806}, 127 (2018)
  doi:10.1007/JHEP06(2018)127
  [arXiv:1804.01939 [hep-ex]].
 
\bibitem{Kang:2018jem} 
  S.~K.~Kang, Z.~Qian, J.~Song and Y.~W.~Yoon,
  arXiv:1810.05229 [hep-ph].
 
\bibitem{Aaboud:2017rel} 
  M.~Aaboud {\it et al.} [ATLAS Collaboration],
  Eur.\ Phys.\ J.\ C {\bf 78}, no. 4, 293 (2018)
  doi:10.1140/epjc/s10052-018-5686-3
  [arXiv:1712.06386 [hep-ex]].

\bibitem{Kauer:2012hd} 
  N.~Kauer and G.~Passarino,
  JHEP {\bf 1208}, 116 (2012)
  doi:10.1007/JHEP08(2012)116
  [arXiv:1206.4803 [hep-ph]
 
\bibitem{Khachatryan:2014iha} 
  V.~Khachatryan {\it et al.} [CMS Collaboration],
  Phys.\ Lett.\ B {\bf 736}, 64 (2014)
  doi:10.1016/j.physletb.2014.06.077
  [arXiv:1405.3455 [hep-ex]].
  
\bibitem{Aaboud:2018puo} 
  M.~Aaboud {\it et al.} [ATLAS Collaboration],
  Phys.\ Lett.\ B {\bf 786}, 223 (2018)
  doi:10.1016/j.physletb.2018.09.048
  [arXiv:1808.01191 [hep-ex]].
  
\bibitem{Alwall:2014hca} 
  J.~Alwall {\it et al.},
  JHEP {\bf 1407}, 079 (2014)
  doi:10.1007/JHEP07(2014)079
  [arXiv:1405.0301 [hep-ph]].
 
\bibitem{Caola:2015psa} 
  F.~Caola, K.~Melnikov, R.~Rontsch and L.~Tancredi,
  Phys.\ Rev.\ D {\bf 92}, no. 9, 094028 (2015)
  doi:10.1103/PhysRevD.92.094028
  [arXiv:1509.06734 [hep-ph]].
  
\bibitem{Cascioli:2014yka} 
  F.~Cascioli {\it et al.},
  Phys.\ Lett.\ B {\bf 735}, 311 (2014)
  doi:10.1016/j.physletb.2014.06.056
  [arXiv:1405.2219 [hep-ph]].
  
  \bibitem{Avery:2012um} 
  P.~Avery {\it et al.},
  Phys.\ Rev.\ D {\bf 87}, no. 5, 055006 (2013)
  doi:10.1103/PhysRevD.87.055006
  [arXiv:1210.0896 [hep-ph]].
 
\bibitem{deFavereau:2013fsa} 
  J.~de Favereau {\it et al.} [DELPHES 3 Collaboration],
  JHEP {\bf 1402}, 057 (2014)
  doi:10.1007/JHEP02(2014)057
  [arXiv:1307.6346 [hep-ex]].
  
    
\bibitem{Goncalves:2018ptp} 
  D.~Goncalves and J.~Nakamura,
  arXiv:1809.07327 [hep-ph].

\end{thebibliography}
\end{document}